\documentclass[letterpaper, 10 pt, conference]{ieeeconf}  

\IEEEoverridecommandlockouts                              

\usepackage{amsmath,amsfonts}
\usepackage{algorithmic}
\usepackage{algorithm}
\usepackage{hyperref}
\usepackage{array}
\usepackage[caption=false,font=normalsize,labelfont=sf,textfont=sf]{subfig}
 \usepackage[table,xcdraw]{xcolor}
\usepackage{textcomp}
\usepackage{stfloats}
\usepackage{url}
\usepackage{multirow}
\usepackage{verbatim}
\usepackage{graphicx}
\usepackage{cite}
\usepackage{subcaption}
\usepackage{authblk}

\title{\LARGE \bf
CL-Flow:Strengthening the Normalizing Flows by Contrastive Learning for Better Anomaly Detection
}

\author{Shunfeng Wang$^{1}$,  Yueyang Li$^{2}$, Haichi Luo$^{2}$ and Chenyang Bi$^{3}$
\thanks{$^{2}$ Yueyang Li is with Faculty of Jiangsu Provincial Engineering Laboratory of Pattern Recognition and Computational Intelligence,
        Jiangnan University, 1800 Lihu Avenue, Wuxi, 214122, Jiangsu, China
        {\tt\small lyueyang@jiangnan.edu.cn}}%
}

\begin{document}

\maketitle
\thispagestyle{empty}
\pagestyle{empty}

\begin{abstract}

In the anomaly detection field, the scarcity of anomalous samples has directed the current research emphasis towards unsupervised anomaly detection. While these unsupervised anomaly detection methods offer convenience, they also overlook the crucial prior information embedded within anomalous samples. Moreover, among numerous deep learning methods, supervised methods generally exhibit superior performance compared to unsupervised methods. Considering the reasons mentioned above, we propose a self-supervised anomaly detection approach that combines contrastive learning with 2D-Flow to achieve more precise detection outcomes and expedited inference processes. On one hand, we introduce a novel approach to anomaly synthesis, yielding anomalous samples in accordance with authentic industrial scenarios, alongside their surrogate annotations. On the other hand, having obtained a substantial number of anomalous samples, we enhance the 2D-Flow framework by incorporating contrastive learning, leveraging diverse proxy tasks to fine-tune the network. Our approach enables the network to learn more precise mapping relationships from self-generated labels while retaining the lightweight characteristics of the 2D-Flow. Compared to mainstream unsupervised approaches, our self-supervised method demonstrates superior detection accuracy, fewer additional model parameters, and faster inference speed. Furthermore, the entire training and inference process is end-to-end. Our approach showcases new state-of-the-art results, achieving a performance of 99.6\% in image-level AUROC on the MVTecAD dataset and 96.8\% in image-level AUROC on the BTAD dataset.

\end{abstract}

\section{INTRODUCTION}

In the contemporary world, industrial products, ranging from aircraft wings to semiconductor chips, require dependable defect detection to ensure quality. Traditional methods were labor-intensive and inefficient. However, recent advances in computer vision and deep learning have revolutionized industrial defect detection, garnering significant attention in both academia and industry. 

In many production processes, due to the scarcity of anomalous samples and the presence of unpredictable anomalous patterns, most of previous anomaly detection methods \cite{chongjian1,chongjian2,chongjian3,chongjian4draem,flow1_CflowAD,flow2_CSflow,flow3_fastflow,flow4_CAINNFlow,flow5_Semi-Push-Pull,flow6_uflow,flow7_pyramidflow} only modeled normal samples. However, it is worth contemplating whether the absence of anomalous samples and their corresponding labels hinders the model from learning more precise mapping relationships. Recently, some studies\cite{cutpaste,flow5_Semi-Push-Pull} have begun incorporating automatically generated negative samples to enhance model performance. But in the work of \cite{cutpaste}, the incorporation of negative samples was not coupled with more intricate network architectures, and the proxy task configurations were overly simplistic, leading to suboptimal performance. In another study\cite{flow5_Semi-Push-Pull}, researchers need to manually set classification boundaries to guide training, and the two-stage training process introduces inconvenience. While the aforementioned methods have their limitations, they also open up new avenues for unsupervised anomaly detection. We can explore the direction of self-supervised learning by utilizing automatically generated negative samples, allowing the network to acquire more precise mapping relationships.

\begin{figure*}[ht]
\centering
\includegraphics[width=1\linewidth]{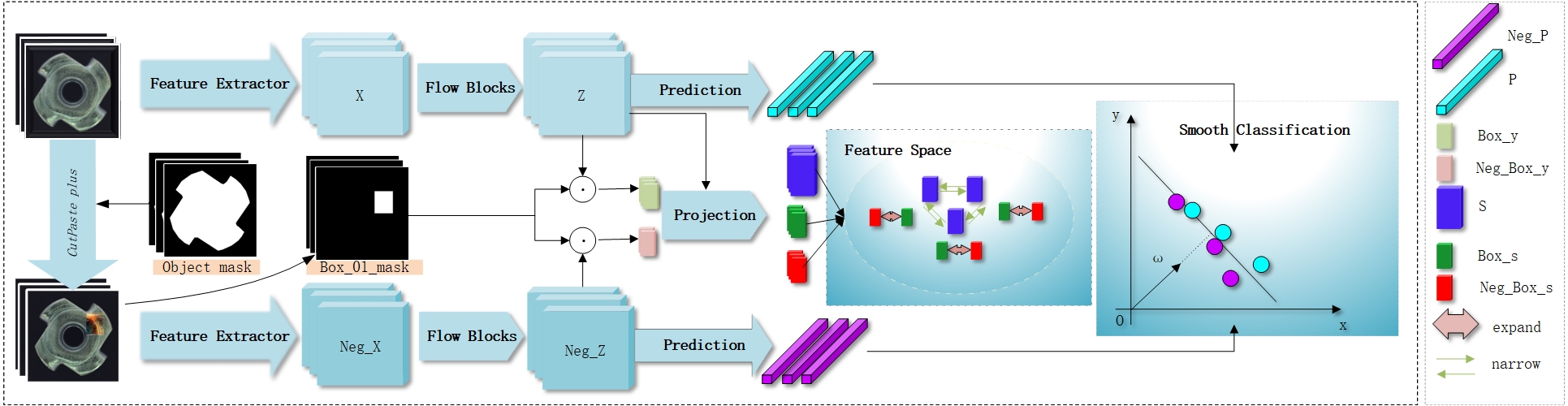}
\caption{Our model framework is an end-to-end Siamese network, where the upper and lower pathways share the same network weights. In the above figure, we employ Cait as the feature extractor, and the Flow Blocks adopt the 2D-Flow module. The Projection module consists of a simple structure comprising linear, ReLU, and linear layers. The Prediction module is a two-layer linear classifier.}
\label{fig_1}
\end{figure*}

In summary, we introduce CL-Flow shown in Fig.\ref{fig_1}, a novel self-supervised anomaly detection framework, which leverages 2D-Flow\cite{flow3_fastflow} as its baseline and enhances model performance through the integration of contrastive learning. Firstly, we proposed an improved version of the CutPaste method to generate realistic abnormal defects and defect masks. Secondly, both positive and negative samples are fed into the network and projected into a new feature space. Finally, distinct feature distance losses and smooth classification losses are employed in the new feature space to optimize network parameters. The detailed description is in sec \ref{sec-cl-flow}.

In addition, we also investigate the widespread applicability of the CL-Flow training approach. Experimental results demonstrate that our method not only enhances the performance of flows models but also shows improvements in the performance of convolutional neural networks\cite{resnet} and transformers\cite{vit,swintransform}. The detailed description is in sec \ref{ssec-fea-ext}.

Overall, our main contributions are as follows:
\begin{itemize}
\item We introduce a novel self-supervised anomaly detection framework with end-to-end training and inference, achieving elevated detection accuracy while retaining the rapid and lightweight characteristics of the 2D-Flow model.
\item We introduce a novel anomaly generation method capable of generating high-quality anomaly samples along with their corresponding region labels. We further incorporate contrastive learning into the 2D-Flow framework to enhance the performance of the base network.
\item We have investigated the inherent limitations of pre-trained models trained on ImageNet\cite{imagenet} when applied to anomaly detection datasets, as they may possess prior information that is not applicable. By using novel pretraining methods, we can effectively enhance the adaptability of the feature extractor.
\item Our method achieves state-of-the-art image-level and pixel-level accuracy on the MVTecAD \cite{MVTecAD_dataset}dataset and the more challenging BTAD\cite{BTADdataset} dataset.
\end{itemize}

\section{Related work}

\subsection{Unsupervised Anomaly Detection}

Existing unsupervised anomaly detection methods can be primarily categorized into reconstruction-based and feature representation-based approaches. Reconstruction-based methods typically employ image generators such as Generative Adversarial Networks (GANs) \cite{GANs,chongjian_Gans_1,chongjian_gans_2,chongjian3,chongjian_gans_3,chongjian_gans_4} or teacher-student models \cite{Student-teacher1,Student-teacher2} to generate or reconstruct input images. These methods also leverage the differences between input images and their reconstructed counterparts to localize the anomalous regions. However, the assumption underlying reconstruction-based methods is not entirely reliable. Even when trained on normal samples only, the model can still fully reconstruct unseen defects and affect detection accuracy.
\begin{figure}
\centering
\includegraphics[width=0.48\textwidth]{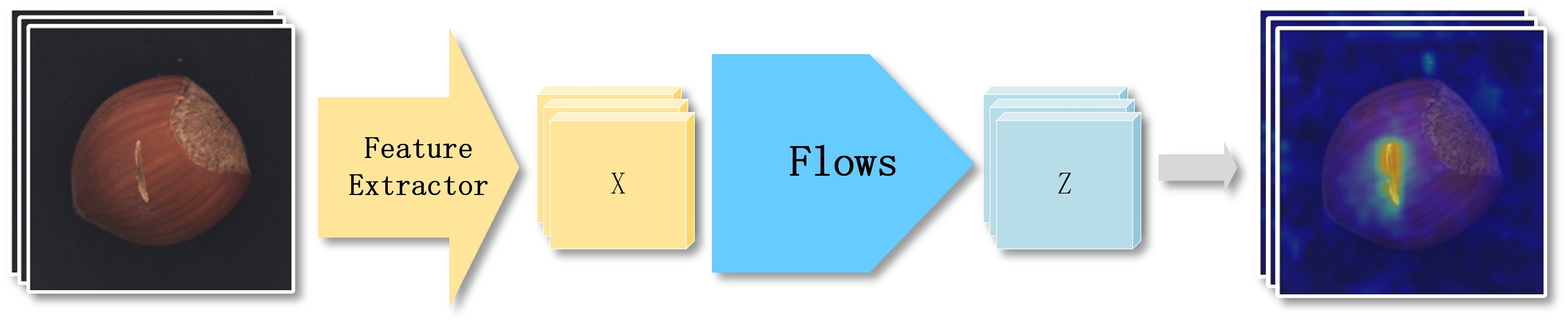}
\caption{The mainstream framework for anomaly detection utilizing normalizing flows.}
\label{fig_2}
\end{figure}

Feature representation-based methods typically utilize pre-trained feature extractors to extract image features. One intuitive approach in this category\cite{feature-distanse1,feature-distanse2,patchcore} is to construct a positive sample database by using normal images as templates and to localize anomalies by measuring the feature differences between negative samples and the positive sample database. However, this approach has the drawbacks of high storage costs and slow retrieval times. Another approach is to use probabilistic modeling to capture the feature distribution of normal samples, which compares the distributions of normal and anomalous regions without additional storage costs. Rippel et al\cite{feature-fenbu1} initially modeled each feature map as a multivariate Gaussian distribution. During testing, the Mahalanobis distance from the feature vector to normal sample distribution detects anomalies. PaDim\cite{feature-fenbu2} further extended the modeling of multivariate Gaussian distributions to the granularity of image blocks, enabling pixel-level segmentation. In contrast, DifferNet\cite{feature-fenbu3} introduced the concept of normalizing flow (NF) in recent years, providing greater potential for such methods.

\subsection{Normalizing Flow}

Normalizing Flows (NF)\cite{NF} transform complex data distributions into simple Gaussian distributions. NF is widely used in generative models and can also perform the inverse transformation. NICE\cite{nice}, RealNVP\cite{Realnvp}, and Glow\cite{glow} are common NF implementations.

Recent works\cite{feature-fenbu3,flow1_CflowAD,flow2_CSflow,flow3_fastflow,flow5_Semi-Push-Pull,flow7_pyramidflow} use pre-trained models to extract features from normal samples and map them to a Gaussian distribution in latent space Z using the NF module. Anomalous samples in Z space exhibit non-Gaussian features with lower likelihood, as shown in Fig.\ref{fig_2}. DifferNet\cite{feature-fenbu3} initially introduced the NF and achieved excellent anomaly classification results at the image level. CFlow-AD\cite{flow1_CflowAD} achieved initial localization of anomalies at the pixel level. FastFlow\cite{flow3_fastflow} introduced the 2D-Flow and accelerated the training and inference process by employing 1x1 convolutions. Although BGAD-FAS\cite{flow5_Semi-Push-Pull} initially used self-generated negative samples to optimize classification boundaries during training, the two-stage process and manual boundary delineation posed inconvenience.

\subsection{Contrastive Learning}
\begin{figure}
\centering
\includegraphics[width=0.48\textwidth]{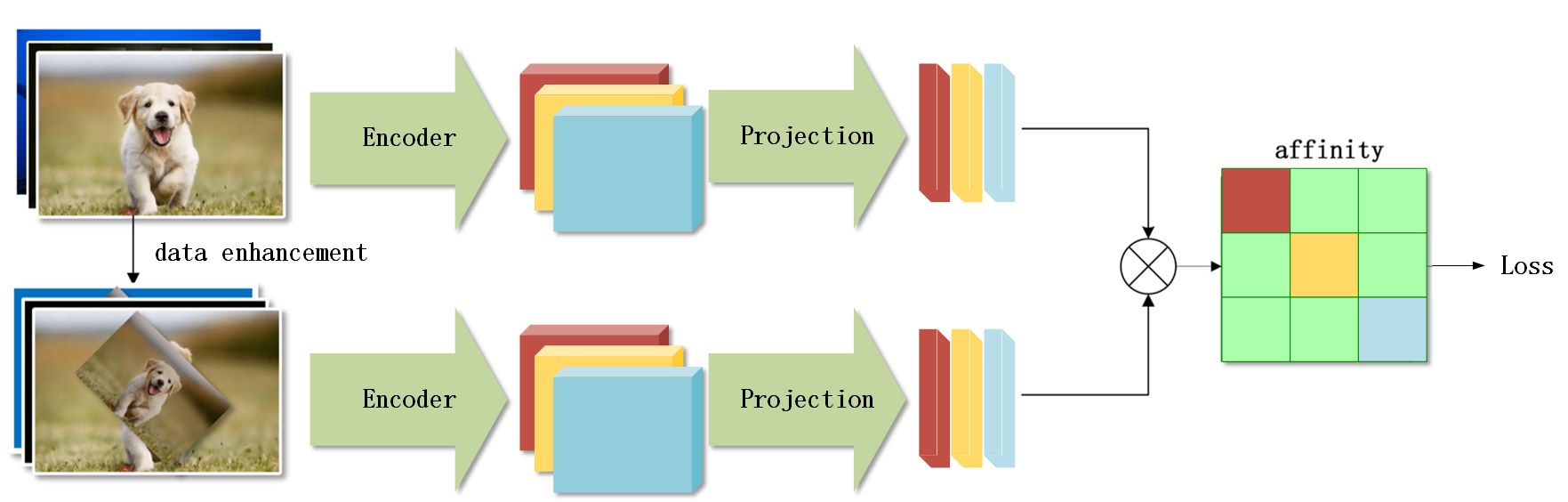}
\caption{The main process of contrastive
learning, which enables the encoder to have excellent feature extraction capabilities.}
\label{fig_3}
\end{figure}

Contrastive learning is a self-supervised learning method used to enhance the feature extraction capability of networks in the absence of labels. It involves setting up proxy tasks and flexible loss functions to train the network.  In recent work, SimCLR\cite{simclr} and MOCO\cite{MOCO1} introduced the concept of automatically generating negative samples and feeding both positive and negative samples into a Siamese network. By maximizing the feature similarity between positive sample pairs within a training batch and minimizing the feature similarity between positive and negative sample pairs, the network learns the features of the dataset, as shown in Fig.\ref{fig_3}.  

Our approach combines contrastive learning with the normalizing flow, introducing a novel self-supervised anomaly detection framework.

\section{Method}

\subsection{CutPaste Plus}
\begin{figure}
\centering
\includegraphics[width=0.35\textwidth,height=1in]{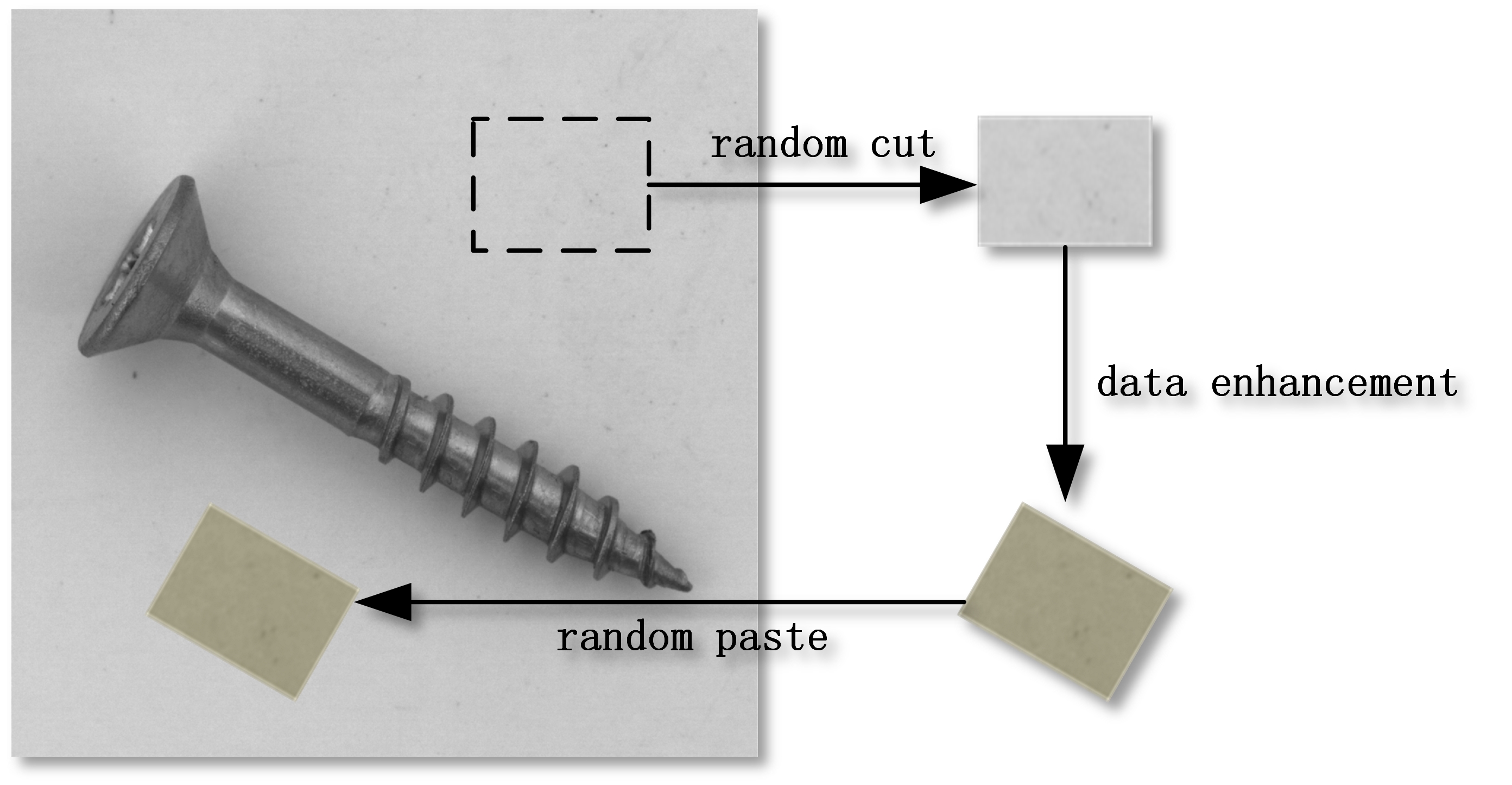}
\caption{The initial CutPaste method generates substandard abnormal images, which undermine the performance of the model. }
\label{fig_4}
\end{figure}
In self-supervised methods that require negative samples, the quality of negative samples is especially crucial. It is essential to emphasize that low-quality negative samples not only fail to augment the model's expressive capabilities but may also inflict substantial detriment to its overall performance.

CutPaste\cite{cutpaste} is a data augmentatio technique that involves randomly cropping patches from the original image and pasting them back into random locations within the same image to create anomaly images. While this method can generate high-quality anomalies in texture datasets, it presents challenges in object datasets. As illustrated in Fig.\ref{fig_4}, these patches are pasted onto the background of the screw images. But in anomaly detection tasks, we often need to disregard the background. Consequently, negative samples produced by this approach are of poor quality and have limited practicality in real-world scenarios. 

In summary, we have made improvements to the CutPaste method. Specifically, we incorporate the FT\cite{FT} saliency detection algorithm to generate object masks for images before training. The FT saliency detection algorithm is a frequency-tuned approach that analyzes an image's energy distribution at various frequencies in order to identify regions of interest. For randomly cropped patches from the original image, we used the object mask to confine their pasting area, ensuring that they solely introduce defects into the foreground of the image. Additionally,  we recorded the defect locations as anomaly region labels, which facilitated subsequent processing of local features in samples. As shown in Table \ref{tab5}, CutPaste Plus significantly enhances anomaly detection across most object-class datasets.
\begin{figure}
\centering
\includegraphics[width=0.48\textwidth]{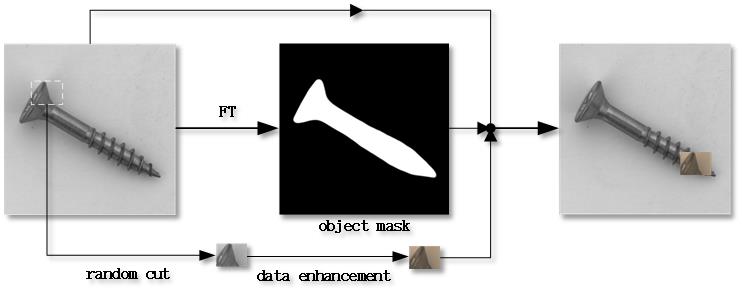}
\caption{The CutPaste Plus method restricts the pasting area of abnormal patches and records it to generate authentic and effective anomaly samples. }
\label{fig_5}
\end{figure}

\subsection{CL-Flow}\label{sec-cl-flow}
In mathematical notation, we represent our normalization flow as $f_\theta: \mathcal{X} \in \mathbb{R}^d \rightarrow \mathcal{Z} \in \mathbb{R}^d$ , where $\mathcal{X} \in \mathbb{R}^d$ denotes the original feature space, $\mathcal{Z} \in \mathbb{R}^d$ denotes the transformed space and $d$ denotes the feature dimensions. In recent studies, the principle of utilizing normalization flows for anomaly detection can be summarized as follows. The normal samples are first passed through a feature extractor, which can be either a VIT or a CNN. The resulting feature maps $x$ are then transformed using the NF into latent tensors $z$ that follow a Gaussian distribution, as shown in the equation:
\begin{equation}
p_X(x)=p_Z(z)\left|\operatorname{det}\left(\frac{\partial z}{\partial x}\right)\right|
\label{equ1}
\end{equation}
We optimize this transformation process by maximizing the log-likelihood:
\begin{equation}
\begin{aligned}
\log p_X(x) & =\log p_Z(z)+\log \left|\operatorname{det}\left(\frac{\partial z}{\partial x}\right)\right| \\
& =\log p_Z\left(f_\theta(x)\right)+\log \left|\operatorname{det}\left(\frac{\partial f_\theta(x)}{\partial x}\right)\right|
\end{aligned}
\label{equ2}
\end{equation}
Here, $\operatorname{det}\left(\frac{\partial f_\theta(x)}{\partial x}\right) $ represents the Jacobian matrix of the invertible transformations ${z}=f_\theta(x)$ and ${x}=f_\theta^{-1}(z)$, and $\theta$ represents the parameters of the flow model. We typically assume $z \sim \mathcal{N}(0, I)$ :
\begin{equation}
p_Z(z)=(2 \pi)^{-\frac{d}{2}} \mathrm{e}^{-\frac{1}{2} z^T z}
\label{equ3}
\end{equation}
Substituting it into equation (2), we have:
\begin{equation}
\begin{aligned}
\log p_X(x)=-\left[
\frac{1}{2} f_\theta(x)^{T} f_\theta(x)-\log \left|\operatorname{det}\left(\frac{\partial z}{\partial x}\right)\right|
+\frac{d}{2} \log (2 \pi)\right]
\end{aligned}
\label{equ4}
\end{equation}

The NF loss function can be defined as:
\begin{equation}
\mathcal{L}_{m l}= \frac{1}{2} f_\theta(x)^{T} f_\theta(x)-\log \left|\operatorname{det}\left(\frac{\partial z}{\partial x}\right)\right|+\frac{d}{2} \log (2 \pi)
\label{equ5}
\end{equation}

In CL-Flow, we continue to use the strategy of NF, but we introduce negative samples to construct a Siamese network and add new loss functions to strengthen the network, as shown in Fig.\ref{fig_1}. During the training phase, the negative samples generated through the CutPaste Plus method are introduced into the network along with the positive samples. Additionally, our CutPaste Plus module records the abnormal region Box\_01\_mask. The resulting Z and Neg\_Z from the Flow Blocks module are multiplied by Box\_01\_mask, yielding Box\_y and Neg\_Box\_y respectively. Subsequently, Box\_y and Neg\_Box\_y, along with Z, are input into the Projection module to obtain Box\_s, Neg\_Box\_s and S. We optimize the network parameters by minimizing the feature distance between the S tensors while expanding the feature distance between Box\_s and Neg\_Box\_s using (\ref{equ6}):

\begin{equation}
\begin{aligned}
\mathcal{L}_{pnd}= & -\log \left\{\sum_{\mathrm{i}=1}^{\mathrm{K}} \sum_{\mathrm{j}=1}^{\mathrm{K}} 1_{[\mathrm{i} \neq \mathrm{j}]} \mathrm{F}\left(\mathrm{S}_{\mathrm{i}}, \mathrm{S}_{\mathrm{j}}\right)+\right. \\
& \left.\sum_{\mathrm{i}=1}^{\mathrm{K}}\left[1-\mathrm{F}\left(\text { Box\_s}_{\mathrm{i}},\text { Neg\_Box\_s}_{\mathrm{i}}\right)\right]\right\}
\end{aligned}
\label{equ6}
\end{equation}
In (\ref{equ6}), $\mathrm{K}$ represents the batch size, $\mathrm{F}$ denotes the cosine similarity function,  $\mathrm{S}$ represents the features of positive samples,  $\mathrm{Neg\_Box\_s}$ represents the features of anomalous regions in negative samples and $\mathrm{Box\_s}$ represents the features of corresponding regions in positive samples. 

Apart from that, Z and Neg\_Z are separately processed through the Prediction module to obtain P and Neg\_P, and further network parameter optimization is achieved using (\ref{equ7}):
\begin{equation}
\begin{aligned}
\mathcal{L}_{s c f}=\text {CELoss }^\tau\left(\mathrm{Neg}\_{\mathrm{P}}, 1\right)+\text { CELoss }{ }^\tau(\mathrm{P}, 0)
\end{aligned}
\label{equ7}
\end{equation}
where $\mathrm{CELoss}$ represents $\mathrm{CrossEntropyLoss}$, and $\tau$ is the label smoothing coefficient.  Table \ref{tab4} have shown that if label smoothing is not applied, $\mathcal{L}_{s c f}$ can negatively impact the model's performance. This is because the object category of anomalous samples is fundamentally the same as that of normal samples, and our goal is to determine whether they contain defects. 

Finally, the overall loss of CL-Flow can be defined as follows:
\begin{equation}
\begin{aligned}
\mathcal{L}=\mathcal{L}_{m l} + \lambda_{1}\mathcal{L}_{p n d} + \lambda_{2}\mathcal{L}_{s c f}
\end{aligned}
\label{equ8}
\end{equation}
During the inference stage, only $\mathcal{L}_{ml}$ is retained for anomaly assessment. The settings of  $\lambda_{1}$ and $\lambda_{2}$ need to ensure that all three loss functions remain on the same scale of magnitude at the beginning of training.
\begin{table*}[ht]
\caption{Complexity Analysis of Anomaly Detection Methods based on Flow. }
\label{tab1}
\centering
\resizebox{0.8\textwidth}{!}{
\begin{tabular}{lcccccccc}
\hline
\multicolumn{1}{c}{}                           & \multicolumn{4}{c}{FPS}                                                                                                                    & \multicolumn{4}{c}{Additional Params}                                                                                                           \\ \cline{2-9} 
\multicolumn{1}{c}{\multirow{-2}{*}{backbone}} & CFlow                        & FastFlow                     & U-Flow                               & CL-Flow(ours)                         & CFlow                         & FastFlow                      & U-Flow                                 & CL-Flow(ours)                          \\ \hline
ResNet18                                       & 40.5 & 61.3 & 32.7         & \textbf{65.2} & 5.5M  & 4.9M  & \textbf{4.3M}  & 4.5M           \\
WideResnet50-2                                 & 24.3 & 40.6 & 21.4         &\textbf{42.1} & 81.6M & 41.3M & \textbf{34.8M} & 36.2M          \\
DeiT-base-distilled                            & 30.0 & 52.1 & -            & \textbf{56.3} & 10.5M & 14.8M & -              & \textbf{10.2M} \\
CaiT-M48-distilled                             & 5.5  & 8.1  & -            & \textbf{8.5}  & 10.5M & 14.8M &-              & \textbf{10.2M} \\
MS-CaIT                                        &-    &-    & \textbf{3.3} &-             & -     & -     & \textbf{12.2M} & -              \\ \hline
\end{tabular}}

\end{table*}
\subsection{Feature Extractor}\label{ssec-fea-ext}

In previous works, the feature extractors were typically pretrained on the ImageNet\cite{imagenet} dataset, which may not be applicable to anomaly detection datasets. Therefore, we investigated novel pretraining approaches on the DeiT\cite{deit} and ResNet\cite{resnet} models, encompassing transfer learning, traditional contrastive learning, as well as our CL-Flow method illustrated in Fig.\ref{fig_1}.

For transfer learning, we appended a classification head to the feature extractor and trained it on the normal samples of the MVTec AD\cite{MVTecAD_dataset} dataset for a 15-class classification task.  And for contrastive learning, we employed SimSiam\cite{simsiam} for transfer learning on the backbone network, which is a traditional contrastive learning method proposed in 2021 and has outperformed contemporary contrastive learning models. 

Furthermore, we employed the CL-Flow approach to train individual feature extractors for each specific class dataset. Specifically, we utilized CutPaste Plus to generate negative samples. For the positive and negative samples' features obtained from the backbone output, we applied cosine similarity loss on the anomaly box regions and used label smoothing for the final binary classification task. The experimental results are presented in Table \ref{tab4}. 

\begin{table*}[ht]
\caption{We present the results in the format (image-level, pixel-level)
of our method and other state-of-the-art approaches in anomaly detection and localization on the MVTec AD dataset.}
\label{tab2}
\centering
\resizebox{0.97\textwidth}{!}{
\begin{tabular}{lcccccccc}
\hline
\textbf{Category}    & \textbf{Draem\textsuperscript{[2021]}} & \textbf{Padim\textsuperscript{[2021]}} & \textbf{Patch-Core\textsuperscript{[2022]}} & \textbf{CFlow\textsuperscript{[2021]}} & \textbf{CSFlow\textsuperscript{[2022]}} & \textbf{FastFlow\textsuperscript{[2021]}} &\textbf{PyramidFlow\textsuperscript{[2023]}} &\textbf{CL-Flow(ours)\textsuperscript{[2023]}} \\ \hline
\textbf{Carpet}      & (97.0,95.5)    & (-,99.1)       & (98.7,98.9)         & (\textbf{100.0},99.3)   & (99.0,-)        & (\textbf{100.0},\textbf{99.4})      & (  -   ,97.4)         & (\textbf{100.0},\textbf{99.4})           \\
\textbf{Grid}        & (99.9,\textbf{99.7})    & (-,97.3)       & (98.2,98.7)         & (97.6,99.0)    & (\textbf{100.0},-)       & (99.7,98.3)       &(      -       ,95.7)          & (\textbf{100.0},98.7)           \\
\textbf{Leather}     & (\textbf{100.0},98.6)   & (-,99.2)       & (\textbf{100.0},99.3)        & (97.7,\textbf{99.7})    & (\textbf{100.0},-)       & (\textbf{100.0},99.5)      & (    -   ,98.7)           & (\textbf{100.0},99.6)           \\
\textbf{Tile}        & (99.6,\textbf{99.2})    & (-,94.1)       & (98.7,95.6)         & (98.7,98.0)    & (\textbf{100.0},-)       & (\textbf{100.0},96.3)      &(   -   ,97.1)           & (\textbf{100.0},96.9)           \\
\textbf{Wood}        & (99.1,96.4)    & (-,94.9)       & (99.2,95.0)         & (99.6,96.7)    & (\textbf{100.0},-)       & (\textbf{100.0},\textbf{97.0})      & (  -   ,\textbf{97.0})           & (\textbf{100.0},96.8)           \\
\textbf{Av. texture} & (99.1,97.8)    & (-,96.9)       & (98.9,95.7)         & (98.7,\textbf{98.5})    & (99.8,-)        & (99.9,98.1)       & (    -   ,97.1)             & (\textbf{100.0},98.3)           \\ \hline
\textbf{Bottle}      & (99.2,\textbf{99.1})    & (-,98.3)       & (\textbf{100.0},98.6)        & (\textbf{100.0},99.0)   & (99.8,-)        & (\textbf{100.0},97.7)      & ( -   ,97.8)             & (\textbf{100.0},98.3)           \\
\textbf{Cable}       & (91.8,94.7)    & (-,96.7)       & (99.5,98.4)         & (\textbf{100.0},97.6)   & (99.1,-)        & (\textbf{100.0},98.4)      & (   -   ,91.8)            & (99.0,\textbf{98.5})            \\
\textbf{Capsule}     & (98.5,94.3)    & (- ,98.5)      & (98.1,98.8)         & (99.3 ,99.0)   & (97.1,-)        & (\textbf{100.0},\textbf{99.1})      & (   -   ,98.6)            & (98.8,\textbf{99.1})            \\
\textbf{Hazelnut}    & (\textbf{100.0},\textbf{99.7})   & (- ,98.2)      & (\textbf{100.0},98.7)        & (96.8 ,98.9)   & (99.6,-)        & (\textbf{100.0},99.1)      & (    -   ,98.1)              & (99.8,99.3)            \\
\textbf{Metal nut}   & (98.7,\textbf{99.5})    & (-,97.2)       & (\textbf{100.0},98.4)        & (91.9,98.6)    & (99.1,-)        & (\textbf{100.0},98.5)      &  (   -   ,97.2)               & (\textbf{100.0},98.3)           \\
\textbf{Pill}        & (98.9,97.6)    & (-,95.7)       & (96.6,97.1)         & (99.9,99.0)    & (98.6,-)        & (99.4,\textbf{99.2})       & (    -   ,94.2)                 & (98.9,99.1)            \\
\textbf{Screw}       & (93.9,97.6)    & (-,98.5)       & (98.1,99.4)         & (99.7,98.9)    & (97.6,-)        & (97.8,99.4)       & ( -   ,94.6)             & (98.3,\textbf{99.5})            \\
\textbf{Toothbrush}  & (\textbf{100.0},98.1)   & (-,98.8)       & (\textbf{100.0},98.7)        & (95.2,99.0)    & (91.9,-)        & (94.4 ,98.9)      & (   -   ,98.5)        & (\textbf{100.0},\textbf{99.2})           \\
\textbf{Transistor}  & (93.1,90.9)    & (-,97.5)       & (\textbf{100.0},96.3)        & (99.1 ,\textbf{98.0})   & (99.3,-)        & (99.8,97.3)       & (   -   ,96.9)            & (\textbf{100.0},97.0)           \\
\textbf{Zipper}      & (\textbf{100.0},98.8)   & (-,98.5)       & (98.8,98.8)         & (98.5 ,\textbf{99.1})   & (99.7,-)        & (99.5,98.7)       & (   -   ,96.6)            & (\textbf{100.0},98.7)           \\
\textbf{Av. objects} & (97.1,97.0)    & (-,97.7)       & (99.1,\textbf{99.3})         & (98.0,98.6)    & (98.2,-)        & (99.1,98.7)       & (    -   ,97.0)             & (\textbf{99.5},98.7)            \\ \hline
\textbf{Av. total}   & (98.0,97.3)    & (97.9,97.5)    & (99.1,98.1)         & (98.3,\textbf{98.6})    & (98.7,-)        & (99.4,98.5)       & (    -   ,97.1)             & (\textbf{99.6},\textbf{98.6})            \\ \hline
\end{tabular}}

\end{table*}

\section{Experiments}
\subsection{Datasets and Metrics}

We evaluated our method on the MVTec Anomaly Detection Dataset\cite{MVTecAD_dataset} (MVTecAD) and BeanTech Anomaly Detection Dataset\cite{BTADdataset} (BTAD). MVTec AD contains high-resolution images from 15 categories with both normal and anomalous samples. The training set only includes normal samples, while the test set contains both. BTAD dataset consists of texture images captured in real industrial scenes, featuring three different categories. The defects present in each category are subtle and difficult to distinguish with the naked eye, posing greater challenges for anomaly classification and localization.

We employed AUROC (Area Under the Receiver Operating Characteristic curve) as the evaluation metric for our experiments. It is widely used in anomaly detection to compare and assess the performance of different models or algorithms. It measures the performance of a classifier by plotting the true positive rate against the false positive rate at various classification thresholds. The AUROC score represents the area under this curve, which ranges from 0 to 1. 
\begin{table}[ht]
\caption{Quantitative results in the format (image-level, pixel-level) of CL-Flow and other outstanding methods approaches in anomaly detection and localization on the BTAD dataset.}
\label{tab3}
\centering
\resizebox{0.48\textwidth}{!}{
\begin{tabular}{ccccc}
\hline
\multirow{2}{*}{\textbf{Methods}} & \multicolumn{3}{c}{\textbf{Classes}}     & \multirow{2}{*}{\textbf{Mean}} \\ \cline{2-4}
                                  & 01           & 02          & 03          &                                \\ \hline
VT-ADL\textsuperscript{[2021]}                            & (97.6,\textbf{99.0})  & (71.0,94.0) & (82.6,77.0) & (83.7,90.0)                    \\
P-SVDD\textsuperscript{[2020]}                            & (95.7,91.6)  & (72.1,93.6) & (82.1,91.0) & (83.3,92.1)                    \\
SPADE\textsuperscript{[2020]}                             & (91.4,97.3)  & (71.4,94.4) & (\textbf{99.9},99.1) & (87.6,96.9)                    \\
PaDiM\textsuperscript{[2021]}                             & (99.8,97.0)  & (82.0,96.0) & (99.4,98.8) & (93.7,97.3)                    \\
PatchCore\textsuperscript{[2022]}                         & (90.9,95.5)  & (79.3,94.7) & (99.8,\textbf{99.3}) & (90.0,96.5)                    \\
FastFlow\textsuperscript{[2021]}                          & (-,95.0)     & (-,96.0)    & (-,99.0)    & (-,96.6)                       \\
PyramidFlow\textsuperscript{[2023]}                       & (\textbf{100.0},97.4) & (88.2,\textbf{97.6}) & (99.3,98.1) & (95.8,97.7)                    \\ \hline
CL-Flow(ours)\textsuperscript{[2023]}                     & (\textbf{100.0},\textbf{99.0}) & (\textbf{91.1},95.8) & (99.4,\textbf{99.3}) & (\textbf{96.8},\textbf{98.0})                    \\ \hline
\end{tabular}}

\end{table}
\begin{table}[h]
\caption{Anomaly detection and localization results in the format (image-level, pixel-level) on the MVTecAD dataset based on CutPaste Plus. 2D-Flow means the baseline network.}
\label{tab4}
\centering
\resizebox{0.48\textwidth}{!}{
\begin{tabular}{ccll}
\hline
                                           & \textbf{w/o Abnormal Samples}        & \multicolumn{2}{c}{\textbf{w/ Abnormal Samples}}                                   \\ \cline{2-4} 
\multirow{-2}{*}{\textbf{Object Category}} & \textbf{2D-Flow}                    & \multicolumn{1}{c}{\textbf{CutPaste}} & \multicolumn{1}{c}{\textbf{CutPaste Plus}} \\ \hline
Bottle                                     & \cellcolor[HTML]{EFEFEF}(\textbf{100.0},97.7) & \cellcolor[HTML]{FBFBDB}(\textbf{100.0},97.6)  & \cellcolor[HTML]{FBFBDB}(\textbf{100.0},\textbf{98.3})       \\
Cable                                      & \cellcolor[HTML]{EFEFEF}(98.5,98.3)  & \cellcolor[HTML]{FBFBDB}(96.5,98.2)   & \cellcolor[HTML]{FBFBDB}(\textbf{99.0},\textbf{98.5})        \\
Capsule                                    & \cellcolor[HTML]{EFEFEF}(98.3,99)    & \cellcolor[HTML]{FBFBDB}(98.5,98.8)   & \cellcolor[HTML]{FBFBDB}(\textbf{98.8},\textbf{99.1})        \\
Hazelnut                                   & \cellcolor[HTML]{EFEFEF}(\textbf{99.9},99.1)  & \cellcolor[HTML]{FBFBDB}(98.3,99.2)   & \cellcolor[HTML]{FBFBDB}(99.8,\textbf{99.3})        \\
Metal nut                                  & \cellcolor[HTML]{EFEFEF}(\textbf{100.0},\textbf{98.5}) & \cellcolor[HTML]{FBFBDB}(\textbf{100.0},97.6)    & \cellcolor[HTML]{FBFBDB}(\textbf{100.0},98.3)       \\
Pill                                       & \cellcolor[HTML]{EFEFEF}(97.6,98.8)  & \cellcolor[HTML]{FBFBDB}(98.3,99.0)   & \cellcolor[HTML]{FBFBDB}(\textbf{98.9},\textbf{99.1})        \\
Screw                                      & \cellcolor[HTML]{EFEFEF}(94.8,99.4)  & \cellcolor[HTML]{FBFBDB}(93.9,99.4)   & \cellcolor[HTML]{FBFBDB}(\textbf{98.3},\textbf{99.5})        \\
Toothbrush                                 & \cellcolor[HTML]{EFEFEF}(94.9,98.9)  & \cellcolor[HTML]{FBFBDB}(91.6,99.0)   & \cellcolor[HTML]{FBFBDB}(\textbf{100.0},\textbf{99.2})       \\
Transistor                                 & \cellcolor[HTML]{EFEFEF}(99.7,96.6)  & \cellcolor[HTML]{FBFBDB}(97.4,\textbf{98.3})   & \cellcolor[HTML]{FBFBDB}(\textbf{100.0},97.0)       \\
Zipper                                     & \cellcolor[HTML]{EFEFEF}(99.5,98.6)  & \cellcolor[HTML]{FBFBDB}(\textbf{100.0},\textbf{98.8})  & \cellcolor[HTML]{FBFBDB}(\textbf{100.0},98.7)       \\ \hline
Mean                                    & \cellcolor[HTML]{EFEFEF}(98.3,98.4)  & \cellcolor[HTML]{FBFBDB}(97.4,98.5)  & \cellcolor[HTML]{FBFBDB}(\textbf{99.5},\textbf{98.7})         \\ \hline
\end{tabular}}

\end{table}
\subsection{ Complexity Analysis}

We evaluated the computational complexity of CL-Flow and other Flow methods in terms of FPS and additional model parameters (excluding the backbone network parameters). The hardware configuration of the machine used for testing is Intel(R) Xeon(R) Silver 4310 CPU @ 2.10GHz and NVIDIA Geforce RTX 3090 GPU (24GB graphics memory). The backbone used in our study were VIT (CaiT and DeiT) and ResNet.  Compared to CFlow, CL-Flow achieves faster inference speed and smaller model parameters due to the utilization of lightweight 2D-Flow modules. In comparison to FastFlow, CL-Flow utilizes fewer (16 blocks for VIT) Flow blocks. When compared to U-Flow, CL-Flow does not require multiple backbone networks applied at different scales, resulting in faster inference speed. The comparison results are shown in Table \ref{tab1}. 

\begin{table*}[ht]
\caption{Detailed investigation experiments on the CL-Flow architecture, including the use of negative samples, the inclusion of projection, and the determination of smoothing coefficient for classification task.
}
\label{tab5}
\centering
\resizebox{0.8\textwidth}{!}{
\begin{tabular}{c|c|cccccccll}
\hline
\multirow{3}{*}{\textbf{Category}} & \textbf{w/o AS}              & \multicolumn{9}{c}{\textbf{w/ AS}}                                                                                                                                                           \\ \cline{2-11} 
                                   & \multirow{2}{*}{\textbf{$\mathcal{L}_{m l}$}} & \multicolumn{2}{c|}{\textbf{$\mathcal{L}_{m l}+\mathcal{L}_{pnd}$}}                            & \multicolumn{4}{c|}{\textbf{$\mathcal{L}_{m l}+\mathcal{L}_{scf}$}}                                                 & \multicolumn{3}{c}{\textbf{$\mathcal{L}_{m l}+\mathcal{L}_{pnd}+\mathcal{L}_{scf}$}} \\ \cline{3-11} 
                                   &                              & w/o projection & \multicolumn{1}{c|}{\textbf{A.w/ projection}} & $\tau$=0          & $\tau$=0.2        & \textbf{B.$\tau$=0.35} & \multicolumn{1}{c|}{$\tau$=0.5}        & \multicolumn{3}{c}{\textbf{A+B}}      \\ \hline
Carpet                             & (99.9,98.6)                  & (99.6,98.9)    & \multicolumn{1}{c|}{(100.0,99.1)}             & (99.9.98.6)  & (98.9,99.2)  & (100.0,99.3)      & \multicolumn{1}{c|}{(98.9,99.1)}  & \multicolumn{3}{c}{(100.0,99.4)}      \\
Grid                               & (100.0,95.2)                 & (99.1,97.1)    & \multicolumn{1}{c|}{(99.9,97.4)}              & (100.0,95.2) & (100.0,98.0) & (100.0,98.2)      & \multicolumn{1}{c|}{(100.0,97.6)} & \multicolumn{3}{c}{(100.0,98.7)}      \\
Leather                            & (100.0,99.3)                 & (100.0,99.5)   & \multicolumn{1}{c|}{(100.0,99.5)}             & (100.0,99.3) & (100.0,99.3) & (100.0,99.3)      & \multicolumn{1}{c|}{(100.0,99.3)} & \multicolumn{3}{c}{(100.0,99.6)}      \\
Tile                               & (99.9,94.4)                  & (100.0,95.3)   & \multicolumn{1}{c|}{(100.0,96.7)}             & (99.9,94.4)  & (100.0,95.4) & (100.0,96.3)      & \multicolumn{1}{c|}{(100.0,95.4)} & \multicolumn{3}{c}{(100.0,96.9)}      \\
Wood                               & (98.4,93.6)                  & (99.4,95.0)    & \multicolumn{1}{c|}{(99.6,96.3)}              & (98.4,93.6)  & (99.5,93.4)  & (100.0,96.8)      & \multicolumn{1}{c|}{(99.7,93.0)}  & \multicolumn{3}{c}{(100.0,96.8)}       \\ \hline
Bottle                             & (100.0,97.7)                 & (100.0,97.8)   & \multicolumn{1}{c|}{(100.0,98.1)}             & (99.8,96.4)  & (99.9,98.1)  & (100.0,98.3)      & \multicolumn{1}{c|}{(100.0,98.0)} & \multicolumn{3}{c}{(100.0,98.3)}      \\
Cable                              & (98.5,98.3)                  & (98.5,98.0)    & \multicolumn{1}{c|}{(99.3,98.4)}              & (88.7,97.3)  & (96.4,97.8)  & (98.4,98.4)       & \multicolumn{1}{c|}{(97.6,98.2)}  & \multicolumn{3}{c}{(99.0,98.5)}       \\
Capsule                            & (98.3,99)                    & (93.0,98.6)    & \multicolumn{1}{c|}{(98.0,99.1)}              & (93.8,97.9)  & (98.3,98.9)  & (98.5,99.0)       & \multicolumn{1}{c|}{(97.8,98.8)}  & \multicolumn{3}{c}{(98.8,99.1)}       \\
Hazelnut                           & (99.9,99.1)                  & (98.0,98.3)    & \multicolumn{1}{c|}{(99.7,99.4)}              & (95.7,97.5)  & (98.5,99.2)  & (99.5,99.2)       & \multicolumn{1}{c|}{(96.8,99.2)}  & \multicolumn{3}{c}{(99.8,99.3)}       \\
Metal nut                          & (100.0,98.5)                 & (100.0,97.7)   & \multicolumn{1}{c|}{(100.0,98.1)}             & (100.0,95.5) & (100.0,98.0) & (100.0,98.3)      & \multicolumn{1}{c|}{(100.0,97.8)} & \multicolumn{3}{c}{(100.0,98.3)}      \\
Pill                               & (97.6,98.8)                  & (95.5,97.8)    & \multicolumn{1}{c|}{(98.9,99.1)}              & (92.8,96.8)  & (98.9,98.7)  & (98.7,99.2)       & \multicolumn{1}{c|}{(98.3,98.5)}  & \multicolumn{3}{c}{(98.9,99.1)}       \\
Screw                              & (94.8,99.4)                  & (78.7,95.0)    & \multicolumn{1}{c|}{(98.0,99.4)}              & (89.6,97.3)  & (95.1,99.4)  & (96.7,99.2)       & \multicolumn{1}{c|}{(89.3,99.2)}  & \multicolumn{3}{c}{(98.3,99.5)}       \\
Toothbrush                         & (94.9,98.9)                  & (99.4,99.1)    & \multicolumn{1}{c|}{(99.4,99.3)}              & (100.0,98.3) & (96.1,98.9)  & (100.0,99.1)      & \multicolumn{1}{c|}{(90.8,99.1)}  & \multicolumn{3}{c}{(100.0,99.2)}      \\
Transistor                         & (99.7,96.6)                  & (97.9,93.1)    & \multicolumn{1}{c|}{(99.8,97.0)}              & (97.2,93.4)  & (99.0,95.9)  & (99.5,96.6)       & \multicolumn{1}{c|}{(99.1,96.0)}  & \multicolumn{3}{c}{(100.0,97.0)}      \\
Zipper                             & (99.5,98.6)                  & (99.3,98.2)    & \multicolumn{1}{c|}{(99.6,98.5)}              & (100.0,98.6) & (99.9,98.6)  & (100.0,98.7)      & \multicolumn{1}{c|}{(99.8,98.7)}  & \multicolumn{3}{c}{(100.0,98.7)}      \\ \hline
Av. total                          & (98.7,97.6)                  & (97.2,97.2)    & \multicolumn{1}{c|}{(99.5,98.4)}              & (97.0,96.7)  & (98.7,97.9)  & (99.4,98.3)       & \multicolumn{1}{c|}{(97.9,97.9)}  & \multicolumn{3}{c}{(99.6,98.6)}       \\ \hline
\end{tabular}}

\end{table*}
\subsection{Comparison with state-of-art methods}

\subsubsection{MVTecAD}
We compared CF-Flow with other state-of-the-art anomaly detection methods, including reconstruction-based (Draem\cite{chongjian4draem}), feature distance-based (Padim\cite{feature-fenbu2}, PatchCore\cite{patchcore}), and most of the Flow-based (CFlow\cite{flow1_CflowAD}, CSFlow\cite{flow2_CSflow}, FastFlow\cite{flow3_fastflow}) approaches under the metrics of image-level AUC and pixel-level AUC. Detailed comparison results for all categories can be found in Table \ref{tab2}. Among them, CL-Flow surpasses all other methods by achieving an Image-AUROC of \textbf{99.6\%} and a Pixel-AUROC of \textbf{98.6\%}.

\subsubsection{BTAD}
We conducted a comparative analysis of various anomaly detection and localization methods, namely VT-ADL\cite{BTADdataset}, P-SVDD\cite{patch-svdd}, SPADE\cite{feature-distanse2}, PatchCore\cite{patchcore}, PaDiM\cite{feature-fenbu2}, FastFlow\cite{flow3_fastflow}, and PyramidFlow\cite{flow7_pyramidflow} on the BTAD dataset. The corresponding results have been presented in Table \ref{tab3}. Among all the algorithms, CL-Flow demonstrated superior performance with a image-level AUROC of \textbf{96.8\%} and a pixel-level AUROC of \textbf{98.0\%}.

\subsection{Ablation Study}

We initially assessed CutPaste Plus against the original CutPaste on object-class datasets, as shown in Table \ref{tab4}. Training with the architecture in Fig.\ref{fig_1}, experiments indicate that inappropriate anomaly generation methods, such as CutPaste, can harm model performance. 

Subsequently, we conducted ablation experiments on the architecture of CL-Flow. In Table\ref{tab5}, \textbf{w/o AS} represents the baseline network without using negative samples, and $\mathcal{L}_{m l}$ refers to the loss function of the baseline network in (\ref{equ8}). \textbf{A} denotes the incorporation of negative samples into $\mathcal{L}_{pnd}$ calculation, which requires feature projection onto a distinct feature space. \textbf{B} represents the optimal performance achieved by adding the $\mathcal{L}_{scf}$ (\ref{equ7}) on the baseline, with a label smoothing coefficient of \textbf{0.35}.

\subsection{Feature extractor and new transfer learning method}

We investigated the impact of different pretraining methods on anomaly detection performance using various feature extractors, including wide\_resnet50\_2\cite{resnet} and Deit\cite{deit}, within the testing framework of 2D-Flow. Different pretraining methods have been explained in sec \ref{ssec-fea-ext}. In Table\ref{tab6}, \textbf{a} represents pre-training using the ImageNet dataset, \textbf{b} represents transfer learning on a 15-class classification task, \textbf{c} represents transfer learning using the SimSiam framework , and \textbf{d} represents
a new transfer learning method under the CL-Flow architecture. The new transfer learning method demonstrated the best performance in terms of image-level AUROC, further validating the effectiveness and comprehensiveness of the CL-Flow training approach.

\begin{table*}[ht]
\caption{Results in the format (image-level, pixel-level) of anomaly detection and localization using different pre-training methods on the feature extractors. 
}
\label{tab6}
\centering
\resizebox{0.8\textwidth}{!}{
\begin{tabular}{ccccccccc}
\hline
                                    & \multicolumn{4}{c}{\textbf{Deit}}                                                                                                                         & \multicolumn{4}{c}{\textbf{wide\_resnet50\_2}}                                                                                                            \\ \cline{2-9} 
\multirow{-2}{*}{\textbf{Category}} & \textbf{a}                           & \textbf{b}                           & \textbf{c}                           & \textbf{d(ours)}                     & \textbf{a}                           & \textbf{b}                           & \textbf{c}                           & \textbf{d(ours)}                     \\ \hline
Carpet                              & \cellcolor[HTML]{EFEFEF}(\textbf{100.0},\textbf{99.3}) & \cellcolor[HTML]{EFEFEF}(\textbf{100.0},99.0) & \cellcolor[HTML]{EFEFEF}(\textbf{100.0},99.0) & \cellcolor[HTML]{EFEFEF}(\textbf{100.0},\textbf{99.3}) & \cellcolor[HTML]{FBFBDB}(99.3,\textbf{98.6})  & \cellcolor[HTML]{FBFBDB}(97.8,97.5)  & \cellcolor[HTML]{FBFBDB}(98.0,97.6)  & \cellcolor[HTML]{FBFBDB}(\textbf{99.9},98.3)  \\
Grid                                & \cellcolor[HTML]{EFEFEF}(98.5,98.0)  & \cellcolor[HTML]{EFEFEF}(98.1,97.7)  & \cellcolor[HTML]{EFEFEF}(98.5,98.0)  & \cellcolor[HTML]{EFEFEF}(\textbf{100.0},\textbf{98.4}) & \cellcolor[HTML]{FBFBDB}(\textbf{100.0},99.2) & \cellcolor[HTML]{FBFBDB}(\textbf{100.0},99.1) & \cellcolor[HTML]{FBFBDB}(\textbf{100.0},98.9) & \cellcolor[HTML]{FBFBDB}(\textbf{100.0},\textbf{99.3}) \\
Leather                             & \cellcolor[HTML]{EFEFEF}(\textbf{100.0},\textbf{99.4}) & \cellcolor[HTML]{EFEFEF}(\textbf{100.0},99.3) & \cellcolor[HTML]{EFEFEF}(\textbf{100.0},99.3) & \cellcolor[HTML]{EFEFEF}(\textbf{100.0},99.3) & \cellcolor[HTML]{FBFBDB}(\textbf{100.0},99.5) & \cellcolor[HTML]{FBFBDB}(\textbf{100.0},98.7) & \cellcolor[HTML]{FBFBDB}(\textbf{100.0},98.5) & \cellcolor[HTML]{FBFBDB}(\textbf{100.0},\textbf{99.6}) \\
Tile                                & \cellcolor[HTML]{EFEFEF}(\textbf{100.0},96.3) & \cellcolor[HTML]{EFEFEF}(\textbf{100.0},\textbf{96.4}) & \cellcolor[HTML]{EFEFEF}(\textbf{100.0},92.7) & \cellcolor[HTML]{EFEFEF}(\textbf{100.0},\textbf{96.4}) & \cellcolor[HTML]{FBFBDB}(\textbf{100.0},96.4) & \cellcolor[HTML]{FBFBDB}(\textbf{100.0},97.4) & \cellcolor[HTML]{FBFBDB}(\textbf{100.0},97.2) & \cellcolor[HTML]{FBFBDB}(\textbf{100.0},\textbf{97.5}) \\
Wood                                & \cellcolor[HTML]{EFEFEF}(97.2,96.1)  & \cellcolor[HTML]{EFEFEF}(99.4,\textbf{97.2})  & \cellcolor[HTML]{EFEFEF}(99.4,96.5)  & \cellcolor[HTML]{EFEFEF}(\textbf{100.0},96.4) & \cellcolor[HTML]{FBFBDB}(99.4,95.6)  & \cellcolor[HTML]{FBFBDB}(98.4,95.6)  & \cellcolor[HTML]{FBFBDB}(99.1,96.2)  & \cellcolor[HTML]{FBFBDB}(\textbf{100.0},\textbf{96.7}) \\ \hline
Bottle                              & \cellcolor[HTML]{EFEFEF}(98.4,\textbf{98.1})  & \cellcolor[HTML]{EFEFEF}(99.6,97.8)  & \cellcolor[HTML]{EFEFEF}(99.5,97.9)  & \cellcolor[HTML]{EFEFEF}(\textbf{100.0},98.0) & \cellcolor[HTML]{FBFBDB}(\textbf{100.0},98.5) & \cellcolor[HTML]{FBFBDB}(\textbf{100.0},98.4) & \cellcolor[HTML]{FBFBDB}(99.5,96.8)  & \cellcolor[HTML]{FBFBDB}(\textbf{100.0},\textbf{98.8}) \\
Cable                               & \cellcolor[HTML]{EFEFEF}(97.9,97.8)  & \cellcolor[HTML]{EFEFEF}(98.2,97.5)  & \cellcolor[HTML]{EFEFEF}(99.1,96.9)  & \cellcolor[HTML]{EFEFEF}(\textbf{100.0},\textbf{98.1}) & \cellcolor[HTML]{FBFBDB}(98.2,\textbf{97.0})  & \cellcolor[HTML]{FBFBDB}(97.1,96.9)  & \cellcolor[HTML]{FBFBDB}(99.7,95.1)  & \cellcolor[HTML]{FBFBDB}(\textbf{100.0},96.9) \\
Capsule                             & \cellcolor[HTML]{EFEFEF}(97.4,98.4)  & \cellcolor[HTML]{EFEFEF}(98.9,\textbf{98.8})  & \cellcolor[HTML]{EFEFEF}(98.1,98.4)  & \cellcolor[HTML]{EFEFEF}(\textbf{99.0},98.5)  & \cellcolor[HTML]{FBFBDB}(98.2,98.8)  & \cellcolor[HTML]{FBFBDB}(98.5,\textbf{99.0})  & \cellcolor[HTML]{FBFBDB}(98.2,98.6)  & \cellcolor[HTML]{FBFBDB}(\textbf{99.1},\textbf{99.0})  \\
Hazelnut                            & \cellcolor[HTML]{EFEFEF}(98.4,99.1)  & \cellcolor[HTML]{EFEFEF}(99.5,\textbf{99.2})  & \cellcolor[HTML]{EFEFEF}(\textbf{99.6},98.8)  & \cellcolor[HTML]{EFEFEF}(\textbf{99.6},99.0)  & \cellcolor[HTML]{FBFBDB}(\textbf{99.5},\textbf{98.3})  & \cellcolor[HTML]{FBFBDB}(98.2,97.8)  & \cellcolor[HTML]{FBFBDB}(98.5,97.9)  & \cellcolor[HTML]{FBFBDB}(99.3,98.1)  \\
Metal nut                           & \cellcolor[HTML]{EFEFEF}(99.7,\textbf{97.5})  & \cellcolor[HTML]{EFEFEF}(99.1,94.6)  & \cellcolor[HTML]{EFEFEF}(98.6,95.7)  & \cellcolor[HTML]{EFEFEF}(\textbf{100.0},97.3) & \cellcolor[HTML]{FBFBDB}(99.9,\textbf{99.4})  & \cellcolor[HTML]{FBFBDB}(99.9,98.5)  & \cellcolor[HTML]{FBFBDB}(99.7,98.8)  & \cellcolor[HTML]{FBFBDB}(\textbf{100.0},99.0) \\
Pill                                & \cellcolor[HTML]{EFEFEF}(97.4,\textbf{98.8})  & \cellcolor[HTML]{EFEFEF}(91.2,98.5)  & \cellcolor[HTML]{EFEFEF}(95.1,97.9)  & \cellcolor[HTML]{EFEFEF}(\textbf{98.2},98.5)  & \cellcolor[HTML]{FBFBDB}(98.8,97.6)  & \cellcolor[HTML]{FBFBDB}(97.0,97.6)  & \cellcolor[HTML]{FBFBDB}(97.6,97.0)  & \cellcolor[HTML]{FBFBDB}(\textbf{99.2},\textbf{98.3})  \\
Screw                               & \cellcolor[HTML]{EFEFEF}(94.1,99.1)  & \cellcolor[HTML]{EFEFEF}(89.6,\textbf{99.2})  & \cellcolor[HTML]{EFEFEF}(86.3,98.3)  & \cellcolor[HTML]{EFEFEF}(\textbf{97.4},98.9)  & \cellcolor[HTML]{FBFBDB}(90.6,98.0)  & \cellcolor[HTML]{FBFBDB}(88.5,97.4)  & \cellcolor[HTML]{FBFBDB}(89.9,98.4)  & \cellcolor[HTML]{FBFBDB}(\textbf{95.3},\textbf{99.2})  \\
Toothbrush                          & \cellcolor[HTML]{EFEFEF}(93.3,\textbf{98.6})  & \cellcolor[HTML]{EFEFEF}(92.2,98.4)  & \cellcolor[HTML]{EFEFEF}(93.0,98.1)  & \cellcolor[HTML]{EFEFEF}(\textbf{95.1},98.5)  & \cellcolor[HTML]{FBFBDB}(95.5,98.4)  & \cellcolor[HTML]{FBFBDB}(93.0,97.8)  & \cellcolor[HTML]{FBFBDB}(93.6,97.8)  & \cellcolor[HTML]{FBFBDB}(\textbf{98.8},\textbf{98.6})  \\
\multicolumn{1}{l}
{Transistor}      & \cellcolor[HTML]{EFEFEF}(99.5,93.9)  & \cellcolor[HTML]{EFEFEF}(99.9,\textbf{97.0})  & \cellcolor[HTML]{EFEFEF}(99.2,95.3)  & \cellcolor[HTML]{EFEFEF}(\textbf{100.0},95.6) & \cellcolor[HTML]{FBFBDB}(99.6,97.6)  & \cellcolor[HTML]{FBFBDB}(97.0,\textbf{98.3})  & \cellcolor[HTML]{FBFBDB}(97.3,97.9)  & \cellcolor[HTML]{FBFBDB}(\textbf{100.0},97.5) \\
\multicolumn{1}{l}
{Zipper}          & \cellcolor[HTML]{EFEFEF}(\textbf{99.4},\textbf{98.3})  & \cellcolor[HTML]{EFEFEF}(98.1,97.8)  & \cellcolor[HTML]{EFEFEF}(98.9,98.1)  & \cellcolor[HTML]{EFEFEF}(\textbf{99.4},98.0)  & \cellcolor[HTML]{FBFBDB}(99.5,98.6)  & \cellcolor[HTML]{FBFBDB}(98.2,98.0)  & \cellcolor[HTML]{FBFBDB}(99.2,98.5)  & \cellcolor[HTML]{FBFBDB}(\textbf{100.0},\textbf{98.9})  \\ \hline
\multicolumn{1}{l}
{Av. total}       & \cellcolor[HTML]{EFEFEF}(98.1,97.9)  & \cellcolor[HTML]{EFEFEF}(97.6,97.9)  & \cellcolor[HTML]{EFEFEF}(97.7,97.4)  & \cellcolor[HTML]{EFEFEF}(\textbf{99.2},\textbf{98.0})  & \cellcolor[HTML]{FBFBDB}(98.6,98.1)  & \cellcolor[HTML]{FBFBDB}(97.6,97.9)  & \cellcolor[HTML]{FBFBDB}(98.0,97.6)  & \cellcolor[HTML]{FBFBDB}(\textbf{99.4},\textbf{98.3})  \\ \hline
\end{tabular}}

\end{table*}
\subsection{Qualitative Results}
\begin{figure}[t]
\centering
\includegraphics[width=0.95\columnwidth]{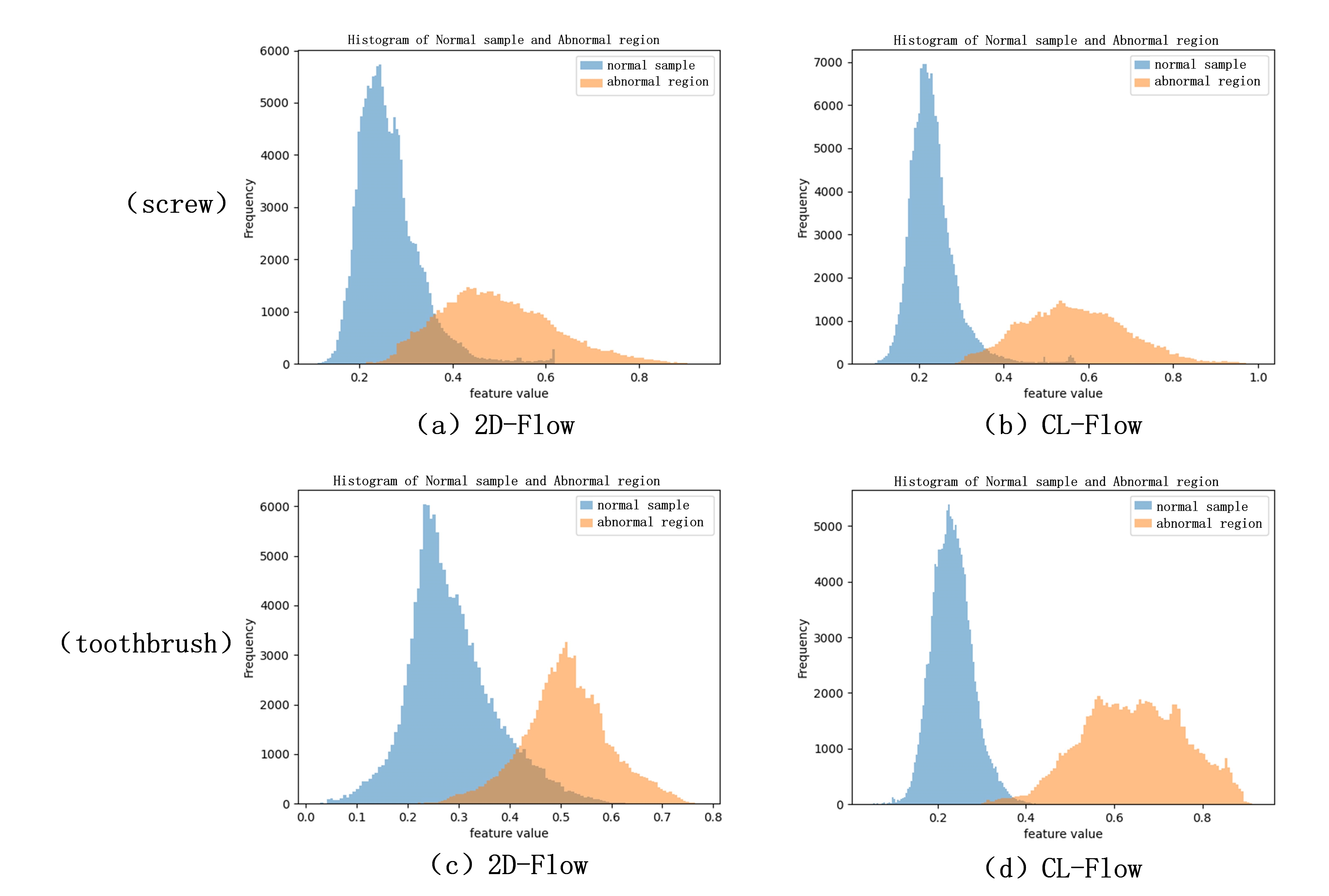}
\caption{Feature distribution plots of the output features from the Flow Blocks. }
\label{fig_6}
\end{figure}

During the inference stage, we utilize Matplotlib to construct output feature distribution maps as shown in Fig.\ref{fig_6}. While CL-Flow does not directly modify the structure of the 2D-Flow module, the employed proxy tasks indirectly optimize the performance of the 2D-Flow module. 

\section{CONCLUSION}
In this paper, we introduce CL-Flow, an efficient self-supervised algorithm for anomaly detection and localization. Our end-to-end approach balances high detection accuracy with the Flow model's speed and lightweight nature. We propose CutPaste Plus for generating realistic negative samples and incorporate contrastive learning into 2D-Flow to optimize model parameters. CL-Flow achieves state-of-the-art image-level and pixel-level accuracy on MVTecAD and BTAD datasets. However, there is potential for improvement, particularly in pixel-level anomaly localization due to the lack of a mainstream multi-scale fusion module. Future work will address this limitation by focusing on fine-grained image feature handling. In summary, CL-Flow presents a promising solution for anomaly detection and localization.





\bibliographystyle{plain}  

\bibliography{ref}

\end{document}